\documentclass[%
 reprint,
superscriptaddress,
 amsmath,amssymb,
 aps,
]{revtex4-1}

\usepackage{braket}
\usepackage{graphicx}
\usepackage{dcolumn}
\usepackage{bm}
\usepackage{comment}
\usepackage{amsthm}
\theoremstyle{definition}

\newtheorem*{theorem*}{Theorem}

\begin{document}

\preprint{APS/123-QED}

\title{High-frequency expansion for Floquet prethermal phases with emergent symmetries: Application to prethermal time crystals and Floquet engineering}

\author{Kaoru Mizuta}
 \email{mizuta.kaoru.65u@st.kyoto-u.ac.jp}
\affiliation{%
 Department of Physics, Kyoto University, Kyoto 606-8502, Japan
}%
\author{Kazuaki Takasan}%
\affiliation{%
 Department of Physics, Kyoto University, Kyoto 606-8502, Japan
}%
\affiliation{%
 Department of Physics, University of California, Berkeley, California 94720, USA
}%
\author{Norio Kawakami}
\affiliation{%
 Department of Physics, Kyoto University, Kyoto 606-8502, Japan
}%

\date{\today}
             

\begin{abstract}
Prethermalization, where quasi-steady states are realized in the intermediate long-time regime (prethermal regime), in periodically driven (Floquet) systems is an important phenomenon since it provides a platform for nontrivial Floquet many-body physics. In this Rapid Communication, we consider Floquet systems with dual energy scales: The Hamiltonian consists of two different terms whose amplitude is either comparable to or much smaller than the frequency. As a result, when the larger-amplitude drive induces a $\mathbb{Z}_N$ symmetry operation, we obtain  the effective static Hamiltonian respecting the emergent $\mathbb{Z}_N$ symmetry in high frequency expansions, which describes the dynamics of such Floquet systems in the prethermal regime. As an application of our formulation, our formalism gives a general way to analyze time crystals in prethermal regimes in terms of the static effective Hamiltonian. We also provide an application to Floquet engineering, with which we can perform the simultaneous control of phases and symmetries of the systems. For example, this enables us to control symmetry protected topological phases even when the original system does not respect the symmetry.
\end{abstract}

\pacs{Valid PACS appear here}
\maketitle



\textit{Introduction.}--- Floquet systems, where the Hamiltonian is periodic in time, have attracted much interest for the past decade since we can control a variety of phases by periodic driving such as laser light \cite{Oka2009,Kitagawa2011,Lindner2011,Grushin2014,Wang2013,Jotzu2014}. It is also remarkable that Floquet systems have novel phases unique to nonequilibrium systems. For example, such unique phases include anomalous Floquet topological insulators (AFTIs) which host chiral edge states despite the vanishing Chern numbers \cite{Kitagawa2010,Rudner2013,Titum2016,Po2016,Mukherjee2017}, and discrete time crystals (DTCs) \cite{Sacha2015,Else2016,Khemani2016}.

One of the most significant phenomena in Floquet systems is prethermalization, in which quasi-steady states are realized in an intermediate time regime. In general, interacting Floquet systems except for many body localized systems \cite{Po2016,Abanin2017} thermalize to a trivial infinite temperature state \cite{Lazarides2014}, thus this intermediate prethermalized regime is quite important for realizing interesting many-body physics in Floquet systems \cite{Takasan2017A,Takasan2017B,Else2017}. In the case where the amplitude of oscillating terms is much smaller than the frequency, prethermalization, called \textit{Floquet prethermalization}, takes place in the intermediate time regime, and then the effective Hamiltonian given by the Floquet Magnus expansion well describes the dynamics \cite{Kuwahara2016,Mori2016,Abanin2017B,Abanin2017Mat}.

However, Floquet systems which realize phases inherent in nonequilibrium inevitably contain a resonant drive, whose amplitude is comparable to the frequency. In this Rapid Communication, we formulate the effective Hamiltonian which well describes Floquet prethermal phases with such resonant drivings, and show that the effective static Hamiltonian, given by the van Vleck expansion, respects an emergent symmetry protected by time translation symmetry. Moreover, we  provide potential applications to time crystals in prethermal regimes and Floquet engineering. We demonstrate a systematic way to analyze the condition for realizing DTC orders in the prethermal regime by the effective Hamiltonian. We also propose symmetry protected Floquet engineering, which allows for the simultaneous control of phases and symmetries by the effective Hamiltonian with an emergent symmetry.

This Rapid Communication is organized as follows. First, we provide the theorem which characterizes Floquet prethermalization under a resonant drive. After that, its application to prethermal time crystals and Floquet engineering is described. We also briefly discuss the control of symmetry protected topological (SPT) phases as a simple example brought by our scheme of Floquet engineering.

\textit{Setup and main results.}--- We consider Floquet systems with a period $T$ on a lattice, that is, we assume that the Hamiltonian satisfies $H(t)=H(t+T)$. Floquet systems can realize various phases unique to nonequilibrium such as anomalous Floquet topological phases and time crystals. However, in such unique Floquet systems, resonant drives are inevitably included, and hence it is difficult to analyze such Floquet systems in a unified and systematic way, in particular, by using standard methods developed for Floquet systems containing only high frequency driving \cite{Mori2016,Kuwahara2016,Abanin2017B,Abanin2017Mat}. Thus, we consider Floquet systems driven by such dual time-dependent terms comparable with the frequency and those much smaller than the frequency, which can realize unique nonequilibrium phases. Here, we obtain the theorem for such Floquet systems, which gives a systematic way to construct the effective static Hamiltonian for a long prethermal regime. The theorem states as follows.
\begin{theorem*}$\,$ \\
Assume that a Floquet system described by 
\begin{equation}\label{Hamiltonian}
H(t)=H_0(t)+V(t),
\end{equation}
where each of $H_0(t)$ and $V(t)$ is a Hamiltonian with a period $T$, satisfies the following conditions.
\begin{flushleft}
(i) The Floquet operator under the Hamiltonian $H_0(t)$ gives an onsite $\mathbb{Z}_N$ symmetry operation, that is,
\begin{equation}
X^N=1, \quad X=\mathcal{T}\exp\left(-i\int^T_0 H_0(t)dt \right).
\end{equation}
\\
(ii) The energy at each site $i$ under the Hamiltonian $V(t)$, which contains at most a $k$-body interaction, is bounded. Then we define the local energy scale of $V(t)$ by $\lambda$.
\end{flushleft}
We define the time evolution operator under $H_0(t),H(t)$ by $U_0(t),U(t)$, respectively, and let $V_m$ be defined as
\begin{equation}\label{VmDef}
V_m = \frac{1}{NT} \int^{NT}_0 U_0^\dagger(t)V(t)U_0(t) e^{im\omega t/N} dt.
\end{equation}
Then, the following statements (a) and (b) are satisfied.
\begin{flushleft}
(a) Let $D_n$, $K_n$ be the effective Hamiltonian and the kick operator of the $n$-th-order truncated van Vleck expansion \cite{Eckardt2015,Bukov2015,Mikami2016,Supplemental}, described by $D_n = \sum_{i=0}^n V_\mathrm{vV}^{(i)}$ and $K_n= \sum_{i=0}^n K^{(i)}$, and the low-order terms are given as follows,
\begin{eqnarray}
V_\mathrm{vV}^{(0)} &=& 0, \: V_\mathrm{vV}^{(1)} = V_0, \:
V_\mathrm{vV}^{(2)} = N\sum_{m\neq0}\frac{[V_{-m},V_m]}{2m\omega}, \\
K^{(0)} &=& 0, \: iK^{(1)} = -N\sum_{m\neq0}\frac{V_m}{m\omega}.
\end{eqnarray}
Then, for any non-negative integer $n$ smaller than $\nu=O(\tilde{\omega})$ (here, $\tilde{\omega}\equiv1/\lambda NT$ is the renormalized frequency), with $U_n(NT) \equiv e^{-iK_n}e^{-iD_n NT}e^{iK_n}$,
\begin{equation}
|| U(NT) - U_n(NT) || = e^{-O(\tilde{\omega})}+O(\tilde{\omega}^{-n-1}), \label{Deviation}
\end{equation}
is satisfied where $\lambda NT$ is small enough compared to $1$.
\\
(b)  For any integer $n$ which satisfies  $0 \leq n < \nu$, the effective Hamiltonian $D_n$ commutes with the symmetry operation $X$, that is,
\begin{equation}
[D_n,X]=0
\end{equation}
is satisfied.
\end{flushleft}
\end{theorem*}

The detailed description of the theorem and its derivation are given in the Supplemental Material \cite{Supplemental}, and hence we provide its physical interpretation here. First, the resonant drive $H_0(t)$, whose energy scale is comparable to the frequency $\omega$, should generate an onsite $\mathbb{Z}_N$ symmetry operation $X$, which acts independently on every site and satisfies $X^N=1$ as the condition (i) (e.g., the Ising symmetry operation is an onsite $\mathbb{Z}_2$ symmetry operation). Such a drive is widely used for realizing nontrivial phases inherent in Floquet systems, such as AFTIs ($X=1$ for a square lattice \cite{Rudner2013, Titum2016}, and $X^2=1$ for a honeycomb lattice \cite{Kitagawa2010,Quelle2017}) and time crystals (shown later). The theorem says that, if the energy scale of $V(t)$ is much smaller than the frequency $\omega$, which means $\lambda NT \ll 1$, the system approximately obeys the time evolution under the Hamiltonian $D_n$, where $n < \nu=O(\tilde{\omega})$ in the coarse-grained stroboscopic dynamics, described as
\begin{equation}
U(NT)\sim e^{-iK_n}e^{-iD_n NT}e^{iK_n}.
\end{equation}
Thus, the state is considered to approach the steady state under the Hamiltonian $D_n$ \cite{Integrability}, and hence we call this regime, dominated by the static Hamiltonian $D_n$, the prethermal regime. The prethermalization itself takes place for an exponentially long time with the frequency $1/\lambda NT$ \cite{Kuwahara2016}, and the time regime, when $D_n$ well describes the dynamics, is long enough as long as the truncation order $n$ satisfies $n< \nu=O(1/\lambda NT)$ \cite{Supplemental}.  In addition, the theorem ensures that $D_n$ possesses the new emergent $\mathbb{Z}_N$ symmetry generated by $X$. It is remarkable that the emergent symmetry of the effective Hamiltonian is completely preserved as long as the time translation symmetry of the original Hamiltonian is conserved. In this sense, the emergent $\mathbb{Z}_N$ symmetry is protected by the time translation symmetry. We also remark that this emergent symmetry is purely generated by the resonant drive $H_0(t)$ and has a significant role in the context of time crystals and Floquet engineering, as discussed later.

We would like to note that a previous study by Else \textit{et al.} also shows the emergence of symmetries in similar Floquet systems \cite{Else2017}. The significant difference is that our  result provides the explicit formulas for the time evolution operator by the van Vleck expansion,  thereby opening up a systematic way for the analysis of DTCs in the prethermal regime including higher-order terms and also another scheme of Floquet engineering \cite{Supplemental}.

\begin{figure}[t]
\begin{center}
    \includegraphics[height=3cm, width=8.5cm]{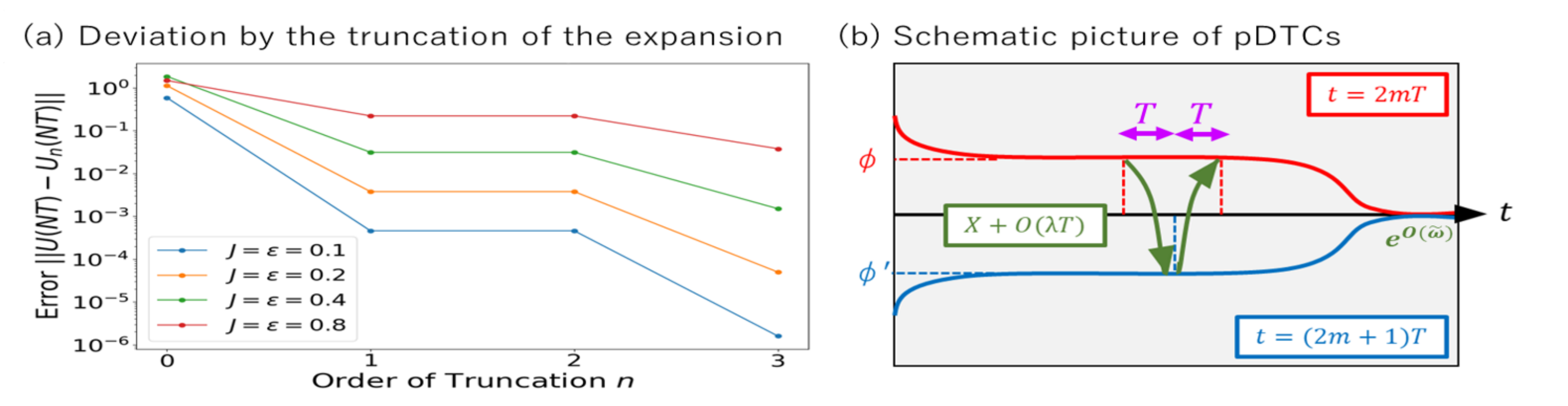}
    \caption{(a) Deviation caused by the truncation of the high-frequency expansion for the model described by Eqs. (\ref{ModelH0}) and (\ref{ModelVt}). The number of the sites $L$ is four. (b) Intuitive picture of pDTCs. Until the prethermal regime ends, DTC orders are possibly realized.}
    \label{DeviationImage}
  \end{center}
\end{figure}

In order to confirm the validity of the theorem, we show a simple example which satisfies the requirement of the theorem. Let us consider a one-dimensional Ising spin system described by
\begin{eqnarray}
H_0(t) &=& \frac{\pi}{2} \sum_i \sigma_i^x \sum_{n \in \mathbb{Z}} \delta(t-nT), \label{ModelH0}\\
V(t) &=& -J\sum_i \sigma_i^z \sigma_{i+1}^z +\varepsilon \sin \omega t \sum_i \sigma_i^x. \label{ModelVt}
\end{eqnarray}
The Hamiltonian $H_0(t)$, which represents a periodic pulse of transverse fields, gives an Ising $\mathbb{Z}_2$ symmetry operation as
\begin{equation}\label{SymmetryOp}
X=\prod_i (-i\sigma_i^x), \qquad X^2=1,
\end{equation}
if the number of the sites is even. The parameter $\lambda$, the energy scale of $V(t)$ at each site, is given by $\lambda \sim \max(J,\varepsilon)$. Here, by the operator norm $||U(NT)-U_n(NT)||$, we numerically evaluate the difference between the time evolution operator $U(NT)$ and the approximated time evolution operator $U_n(NT)$  [see Fig. \ref{DeviationImage} (a)]. As the theorem states in Eq. (\ref{Deviation}), the deviation $||U(NT)-U_n(NT)||$  exponentially decreases with the truncation order $n$ (unless the $n$-th-order term disappears), or decreases as a power of $\lambda NT$.

\textit{DTC orders in the prethermal regime.}--- As an application of the theorem, we show that the theorem confirms the possibilities of DTC orders in the prethermal regime and also gives a systematic way to analyze these time crystalline orders in detail, including their order parameters and the effective temperature. 

DTCs are ordered phases where a discrete time translation symmetry is spontaneously broken \cite{Watanabe2015,Khemani2017,Huang2018}. In other words, in DTCs, certain local observables oscillate with an integer multiple period of the Hamiltonian in the steady state, and the oscillation possesses robustness to perturbations. Ordinary DTCs require randomness which induces many-body localization to avoid the thermalization to infinite temperature \cite{Else2016,Keyserlingk2016,Zhang2017,Yao2017}. However, even in closed Floquet systems without randomness, robust DTC orders can be observed in their quasi-steady states, corresponding to the steady states in the prethermal regime \cite{Else2017,Zeng2017,Mizuta2018}. Here, we call them prethermal discrete time crystals (pDTCs) to distinguish them from ordinary DTCs.

Let us show how the robust DTC orders appear during the prethermal regime in Floquet systems driven by the Hamiltonian Eq. (\ref{Hamiltonian}). For simplicity, we consider an Ising spin system and assume that $X$ defined by Eq. (\ref{SymmetryOp}) corresponds to a $\mathbb{Z}_2$ symmetry operation. Then, by assuming that the energy scale of $V(t)$ is much smaller than $\omega$, the time evolution operator for $2T$ is approximated as $U(2T) \sim e^{-iK_n} e^{-iD_n 2T} e^{iK_n}$.

If we focus on stroboscopic dynamics at $t=2mT$, the state undergoes the time evolution as
\begin{equation}
\ket{\psi(2mT)}\sim e^{-iK_n} e^{-iD_n 2mT} e^{iK_n} \ket{\psi(0)}.
\end{equation}
Since the kick operator $K_n$ only gives a unitary transformation, the system equilibrates under the effective Hamiltonian $D_n$ \cite{Integrability}. Assume that $\sigma_i^z$ is the order parameter for spontaneous breaking of the symmetry represented by $X$ under the Hamiltonian $D_n$. Then, after the equilibration, its expectation value $\bra{\psi(2mT)}\sigma_i^z\ket{\psi(2mT)}$ approximately approaches a certain constant $\phi$ \cite{OrderParameter}. As well, if the stroboscopic dynamics at $t=(2m+1)T$ is focused on, the observable $\bra{\psi((2m+1)T)}\sigma_i^z\ket{\psi((2m+1)T)}$ approaches a certain constant $\phi '$ in the same time scale. With the zero-th-order approximation $U(T)=X+O(\lambda T)$, the order parameters $\phi$ and $\phi '$ are related as follows \cite{ZerothOrder},
\begin{eqnarray}
\phi ' &=& \bra{\psi((2m+1)T)}\sigma_i^z\ket{\psi((2m+1)T)} \nonumber \\
&=& \bra{\psi(2mT)} X^\dagger \sigma_i^z X \ket{\psi(2mT)} +O(\lambda T) \nonumber \\
&=& -\phi + O(\lambda T).
\end{eqnarray}
Thus, unless the order parameter $\phi=O(\lambda T)\ll 1$, the observable $\sigma_i^z$ oscillates with the period $2T$ as shown in Fig. \ref{DeviationImage} (b), which is a hallmark of a DTC order. Here, we have considered the lowest order of the order parameters to confirm the order parameters alternate every period, but once we confirm the existence of DTC orders in the prethermal regime, we can evaluate the order parameters including higher-order correction terms by the effective Hamiltonian $D_n$. 

We can extend this discussion to a general case where $X$ is a $\mathbb{Z}_N$ symmetry operation. In that case, if the $\mathbb{Z}_N$ symmetry is spontaneously broken under the effective Hamiltonian $D_n$, a robust $NT$-periodic oscillation appears in the prethermal time regime. Here, the inverse-temperature of the system $\beta$ is determined by the initial state $\ket{\psi(0)}$, the kick operator $K_n$, and the effective Hamiltonian $D_n$ as
\begin{equation}
\bra{\psi(0)} e^{iK_n} D_n e^{-iK_n} \ket{\psi(0)}= 
\frac{\mathrm{Tr} [D_n e^{-\beta D_n}]}{\mathrm{Tr}[e^{-\beta D_n}]}.
\end{equation} 
This formula represents the energy conservation during the prethermal regime. In other words, if the effective temperature corresponding to the initial state is lower than the critical temperature of $D_n$, a DTC order appears in the prethermal regime. We can also treat the robustness of pDTCs to any perturbations which do not break the time translation symmetry, such as a deviation from the exact spin flip, by including the perturbations into $V(t)$ in Eq. (\ref{Hamiltonian}) and analyzing the new effective Hamiltonian. That is, the robustness of the subharmonic oscillation relies on the long-range order brought on by the spontaneous symmetry breaking under the static Hamiltonian $D_n$.

\begin{figure*}
\hspace{-1cm}
\begin{center}
    \includegraphics[height=4cm, width=18cm]{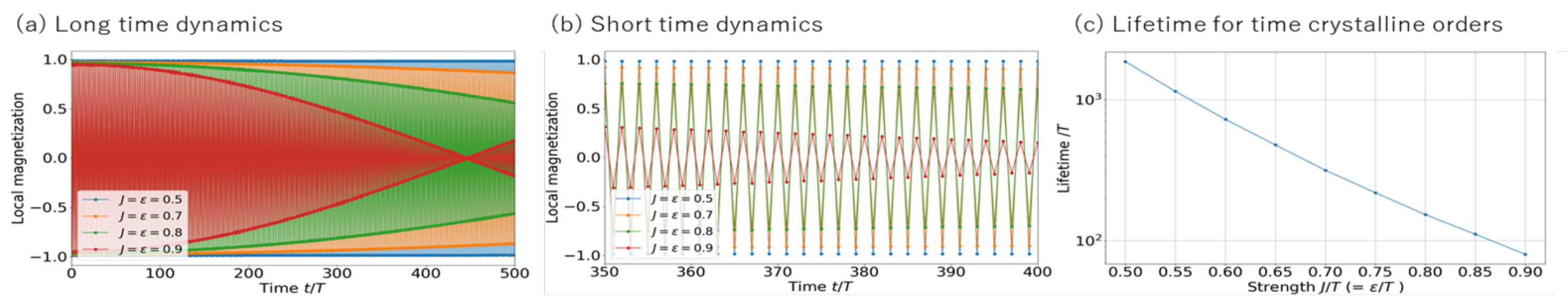}
    \caption{Numerical results for the model described by Eqs. (\ref{ModelH0}) and (\ref{ModelVt}). Each of them is calculated by the exact diagonalization for the finite size $L=4$. (a), (b) Stroboscopic dynamics of the local magnetization $\bra{\psi(t)}\sigma_i^z\ket{\psi(t)}$. The local magnetization oscillates with the period $2T$ in the long transient dynamics. (c) Lifetime of prethermal DTC orders. Here, the lifetime is defined by the time when the amplitude of the oscillation reaches 0.95 times as large as the first one. } 
    \label{TimeCrystalImage}
  \end{center}
\end{figure*}

Let us now provide a simple example and show how we can analyze the pDTC by the effective Hamiltonian and the unitary transformation given by the van Vleck expansion. The model is a driven Ising spin chain described by Eqs. (\ref{ModelH0}) and (\ref{ModelVt}). When we truncate the van Vleck expansion by the first order, we obtain the effective Hamiltonian and the kick operator as follows:
\begin{eqnarray}
D_1 &=& - J \sum_i \sigma_i^z \sigma_{i+1}^z, \\
K_1 &=& - \frac{\varepsilon T}{4\pi} \sum_i \sigma_i^x.
\end{eqnarray}
Thus, since $D_1$ is a one-dimensional Ising Hamiltonian, spontaneous symmetry breaking (SSB) under $D_1$ takes place when the interaction $J$ is nonzero and the unitarily-transformed initial state $e^{-iK_1}\ket{\psi(0)}$ corresponds to a zero temperature state under $D_1$. In principle, symmetry breaking is caused by observation, decoherence, or perturbations which break the underlying symmetry. Here, we prepare the symmetry-broken initial states $e^{iK_1}\ket{\uparrow\uparrow \cdots \uparrow}$ or  $e^{iK_1}\ket{\downarrow\downarrow \cdots \downarrow}$, whose origins can be attributed to the observation. Since the initial states are already the steady states under the Hamiltonian $D_1$, the integrability of the effective Hamiltonian is not important in this case.

Figure \ref{TimeCrystalImage} shows numerical results for the stroboscopic dynamics of the local $z$ spin $\bra{\psi(nT)}\sigma_i^z\ket{\psi(nT)}$ and the lifetime of the oscillation when the initial state $\ket{\psi(0)}$ is $e^{iK_1}\ket{\uparrow\cdots\uparrow}$. We do not see the relaxation since the initial state itself corresponds to the steady state under $D_1$. However, we can estimate the lifetime of the prethermal regime from the time period where $|\sigma_i^z(nT)|$ is nearly $1$ [Fig. \ref{TimeCrystalImage} (a)], since the latter time scale gives the lower bound of the lifetime of the prethermal regime. From the figure, it is confirmed that a robust $2T$-periodic oscillation, which is a sign of pDTCs, is observed within the long transient dynamics when we prepare the initial state predicted by the first order approximation [Fig. \ref{TimeCrystalImage} (b)].  We can also see that the lifetime of the pDTCs exponentially increases with the frequency [Fig. \ref{TimeCrystalImage} (c)].


\textit{Symmetry protected Floquet engineering.}--- Here, we would like to show another application of the theorem to Floquet engineering, which enables the simultaneous control of phases and symmetries of the system. 

Floquet engineering is a scheme to obtain preferable systems or control phases by realizing a proper effective static Hamiltonian under a periodic drive. In conventional ways, the energy scale of the periodic drive is much smaller than the frequency $\omega$, and the effective Hamiltonian is obtained by the high-frequency expansion. High-order correction terms result in an essentially different Hamiltonian, thereby enabling us to control ordered phases or topological phases. This technique for generic Floquet systems is based on the conservation of the effective Hamiltonian given by the high-frequency expansion within the long time regime \cite{Mori2016,Kuwahara2016,Abanin2017B,Abanin2017Mat}.

Similarly, our formalism, which gives the effective Hamiltonian with the emergent symmetry, provides another way of Floquet engineering. Suppose that we drive the system as Eq. (\ref{Hamiltonian}). Then, if we focus on the stroboscopic dynamics at $t=mNT$ ($m \in \mathbb{N}$), the theorem dictates that the effective static system described by the Hamiltonian
\begin{equation}
D_n=V_0+N\sum_{l \neq 0} \frac{[V_{-l},V_l]}{2l\omega}+\lambda O((\lambda T)^2) , \quad XD_n X^{-1}=D_n
\end{equation}
is realized in the prethermal regime. Significantly, we can control phases by high order correction terms, which are essentially different terms originating from the commuters, as with the ordinary Floquet engineering, and moreover, we can add the symmetry represented by $X$ to the system.

There are several attempts to realize emergent symmetries by a periodic drive \cite{Iadecola2015,Potirniche2017}. However, in these studies, only the zero-th-order term acquires an emergent symmetry. Thus, it is worth noting that our proposal gives the effective Hamiltonian which is symmetric including higher-order terms for general Floquet systems.

Finally, we provide the control of SPT phases as a simple example of our scheme of Floquet engineering. Let us consider a one-dimensional spin system with $S=1$ described by
\begin{equation}\label{ModelSPT}
H(t)=\pi \sum_i S_i^x \sum_n \delta(t-nT) + \eta H_\mathrm{AKLT} + \mu V(t).
\end{equation}
The AKLT (Affleck-Kennedy-Lieb-Tasaki) Hamiltonian $H_\mathrm{AKLT}=\sum_i \{ \vec{S}_i \cdot \vec{S}_{i+1} + (\vec{S}_i \cdot \vec{S}_{i+1})^2/3 \}$ is a topologically nontrivial Hamiltonian under the $\mathbb{Z}_2 \times \mathbb{Z}_2$ symmetry ($\pi$-spin rotation around the $x,z$- axes) \cite{Affleck1987,Pollmann2010,Pollmann2012,Tasaki2018}. Here, we assume that $V(t)$ is symmetric under the $\pi$-spin rotation around the $z$- axis, and that the time average of $V(t)$ over one period is zero. In the high-frequency regime $\lambda \equiv\max(\eta,\mu) \ll \omega$, by regarding the first term of Eq. (\ref{ModelSPT}) as $H_0(t)$, we can realize the effective static system whose Hamiltonian is
\begin{equation}
D_2 = \eta H_\mathrm{AKLT} + \frac{\mu^2}{\omega} V^{(2)}, \quad V^{(2)}=\sum_{l \neq 0} \frac{[V_{-l},V_l]}{l},
\end{equation}
where we have performed the truncation at the second order. Importantly, the effective Hamiltonian $D_2$ possesses the emergent $\mathbb{Z}_2 \times \mathbb{Z}_2$ symmetry, which allows nontrivial topological phases. Then we choose $V(t)$ so that $V^{(2)}$ can be a trivial Hamiltonian [e.g., $V^{(2)}$ is proportional to the topologically trivial Hamiltonian $\sum_i \{(S_i^x)^2-(S_i^y)^2\}$, when we choose $V(t)=\sum_i (S_i^x S_i^y e^{i\omega t} + \mathrm{H.c.})$]. By tuning the frequency $\omega$ or the intensity $\mu$, corresponding to the continuous deformation between the nontrivial Hamiltonian and the trivial Hamiltonian, a topological phase transition is expected to be observed, which represents the possibility of controlling SPT phases. It should be noted that we can perform this control of topological phases protected by the $\mathbb{Z}_2 \times \mathbb{Z}_2$ symmetry, though the original system does not respect the $\mathbb{Z}_2 \times \mathbb{Z}_2$ symmetry.


\textit{Discussion and conclusions.}--- In this Rapid Communication, we have studied Floquet systems driven by a resonant drive which induces onsite $\mathbb{Z}_N$ symmetry operations. We have clarified that such Floquet systems are well described by the effective static Hamiltonian given by the van Vleck expansion and acquire an emergent symmetry during the long prethermal regime. Importantly, this property opens up several potential applications that are not accessible by previous studies, that is, it gives a general way to analyze time crystalline orders in the prethermal regime including higher order terms, or it enables simultaneous control of phases and symmetries by a periodic drive.

With our results, we can analytically discuss the phase transition points or scaling laws of prethermal discrete time crystals. Also, it would be interesting to find a way to realize nontrivial SPT phases in systems which host trivial phases in the undriven case with our scheme of Floquet engineering. They are left for future work.

\begin{acknowledgments}
This work is supported by a Grant-in-Aid for Scientific
Research on Innovative Areas ``Topological Materials Science''
(KAKENHI Grant No. JP15H05855) and also JSPS KAKENHI
(Grants No. JP16J05078, No. JP18H01140, and JP19H01838). K.T. thanks JSPS for support from a Research Fellowship for Young Scientists. 
\end{acknowledgments}

\providecommand{\noopsort}[1]{}\providecommand{\singleletter}[1]{#1}%

\clearpage

\newcommand{\wsq}{\qquad $\square$}
\newtheorem*{proof*}{Proof}
\newtheorem{lemma}{Lemma}
\renewcommand{\thesection}{S\arabic{section}}
\renewcommand{\theequation}{S\arabic{equation}}
\setcounter{equation}{0}
\renewcommand{\thefigure}{S\arabic{figure}}
\setcounter{figure}{0}

\onecolumngrid
\begin{center}
 {\large 
 {\bfseries Supplemental Materials for \\ ``High-frequency expansion for Floquet prethermal phases \\ with emergent symmetries: \\ Application to time crystals and Floquet engineering'' }}
\end{center}
\vspace{10pt}

\section{Main theorem and Lifetime of prethermalization}\label{Derivation}

In this section, we describe the theorem depicted in the main text and show its derivation.
\subsection{Theorem}

\begin{theorem*}

Assume that the following conditions (i)---(iii) are satisfied.
\begin{flushleft}
(i) The periodic Hamiltonian $H(t)$ is given by 
\begin{equation}\label{Hamiltonian-S}
H(t)=H_0(t)+V(t),
\end{equation}
and both of $H_0(t)$ and $V(t)$ are periodic Hamiltonians which share the same period $T$.
\\
(ii) The dynamics by the Hamiltonian $H_0(t)$ represents an onsite unitary $\mathbb{Z}_N$ symmetry operation. In other words, $H_0(t)$ can be written in the form of $H_0(t)=\sum_i h_i(t)$, where $h_i(t)$ nontrivially acts only on a site $i$, and 
\begin{equation}\label{SymmetryOp-S}
X^N = 1 \quad \mathrm{where} \quad X = \mathcal{T} \mathrm{exp} \left( -i \int^T_0 H_0(t)dt \right)
\end{equation}
is satisfied.
\\
(iii) The Hamiltonian $V(t)$ contains the at most $k$-body interaction and the energy at each site $i$ is bounded by a certain constant $J$, that is,
\begin{equation}\label{energy_bound}
\sum_X ||v_X(t)|| < J
\end{equation}
is satisfied (here $v_X(t)$ represents a term included in $V(t)$ which nontrivially acts on the site $i$ and the sum is taken over all such operators). We take $\lambda = 2kJ$ as a typical energy scale of the Hamiltonian $V(t)$.
\end{flushleft}

Let us define the time evolution operator of the system by
\begin{equation}
U(t)=\mathcal{T} \mathrm{exp} \left( -i \int^t_0 H(t')dt' \right).
\end{equation}
Then, the following statements are satisfied.
\begin{flushleft}

(a) Let $\nu$ be the largest integer which does not exceed $1/16\lambda NT$. Then, for any integer $n$ which satisfies  $0 \leq n < \nu$, 
\begin{equation}\label{U_NT}
|| U(NT) - e^{-iK_n} e^{-i D_n NT}e^{iK_n} || = e^{-O(\tilde{\omega})} + O(\tilde{\omega}^{n+1})
\end{equation}
is satisfied with $\tilde{\omega}\equiv 1/\lambda NT$ when $\lambda NT < 1/4$.
Here, $D_n$ and $K_n$ are respectively the effective Hamiltonian and the kick operator of the $n$-th order truncated van Vleck expansion given by
\begin{eqnarray}
D_n &=& \sum_{i=0}^n V_\mathrm{vV}^{(i)}, \\
K_n &=& \sum_{i=0}^n K^{(i)}.
\end{eqnarray}
We provide the $i$-th order terms $V_\mathrm{vV}^{(i)}$ and $K^{(i)}$ of the van Vleck expansion in Eqs. (\ref{vanVleck0})---(\ref{kick3}) for a small integer $i$. \\

(b)  For any integer $n$ which satisfies  $0 \leq n < \nu$, the effective Hamiltonian $D_n$ commutes with the symmetry operation $X$, that is,
\begin{equation}
[D_n,X]=0.
\end{equation}
\end{flushleft}
\end{theorem*}

\subsection{Derivation of the theorem}
The proof of the theorem is composed of two steps. The first step is that we apply Kuwahara et al.'s theorem \cite{Kuwahara2016-S} that shows prethermalization under high-frequency driving in the interaction picture. The second one is to find the effective Hamiltonian and the unitary transformation from the Floquet Magnus expansion so that the effective Hamiltonian commutes with the $\mathbb{Z}_N$ symmetry operation $X$. 

First, let us move on to the interaction picture with regarding $H_0(t), V(t)$ as a nonperturbative part, and an interacting part, respectively. Then, the interaction picture of $V(t)$ is defined as follows,

\begin{equation}
V_\mathrm{int}(t) = U_0(t) V(t) U_0^\dagger(t) \quad \mathrm{where}  \quad U_0(t) = \mathcal{T} \mathrm{exp} \left( -i \int^t_0 H_0(t')dt' \right) .
\end{equation}

Then, because of the periodicity of $V(t)$ and the assumption $X^N=1$, $V_\mathrm{int}(t)$ has the period $NT$. In addition, since $H_0(t)$ is composed of onsite operators, $V_\mathrm{int}(t)$ includes the at most $k$-body interaction. With considering the fact that the energy at each site of $V_\mathrm{int}(t)$ is also bounded by $J$ as Eq. (\ref{energy_bound}), we can apply the theorem by Kuwahara et al. \cite{Kuwahara2016-S} to this $NT$-periodic Floquet system described by $V_\mathrm{int}(t)$ when we assume $\lambda NT < 1/4$. Kuwahara's theorem rigorously states that, in Floquet systems whose frequency is much larger than their energy scale, their stroboscopic dynamics can be approximated by the one under the effective Hamiltonian given by the truncated Floquet-Magnus expansion. In other words, if the truncation order $n$ is smaller than $\nu=O(\tilde{\omega})$, where $\nu$ is the largest integer that does not exceeds $1/16\lambda NT$, the Floquet operator $U(NT)$ satisfies
\begin{equation}\label{error}
|| U(NT) - e^{-iV_\mathrm{FM}^n NT} || = e^{-O(\tilde{\omega})} + O(\tilde{\omega}^{-n-1}).
\end{equation}
The effective Hamiltonian $V_\mathrm{FM}^n$ is given by truncation of the Floquet Magnus expansion up to the $n$-th order. The concrete form of the Floquet Magnus expansion is provided by Eqs. (\ref{FloquetMagnus0})---(\ref{FloquetMagnus3}) in the section \textbf{\ref{SecExpansion}}.

In the high frequency regime $\lambda NT \ll 1$, the right hand side of Eq. (\ref{error}) is quite small. This implies that, $U(NT) \sim e^{-iV_\mathrm{FM}^n NT}$, that is, the exact $NT$-stroboscopic dynamics can be approximated by the one under the effective Hamiltonian $V_\mathrm{FM}^n$. However, $V_\mathrm{FM}^n$ given by the truncated Floquet Magnus expansion as Eqs. (\ref{FloquetMagnus0})---(\ref{FloquetMagnus3}) does not necessarily commute with the $\mathbb{Z}_N$ symmetry operation $X$. Thus, the next step is to find  a proper unitary transformation so that the effective Hamiltonian commutes with $X$.

Here, we choose the unitary transformation which transforms the Floquet Magnus expansion to the van Vleck expansion. The  explicit form of the van Vleck expansion is given in Eqs. (\ref{vanVleck0})---(\ref{kick3}). When we truncate these high frequency expansions up to a finite $n$-th order, they are related as follows:
\begin{equation}
|| e^{-iK_n} e^{-iV_\mathrm{vV}^n NT} e^{iK_n} - e^{-iV_\mathrm{FM}^n NT} || = O((\lambda NT)^{n+1}).
\end{equation}
$K_n$, which generates the unitary transformation, is a kick operator truncated up to the $n$-th order, and $V_\mathrm{vV}^n$ is the van Vleck effective Hamiltonian truncated up to  the $n$-th order. We thus end up with 
\begin{eqnarray}
|| U(NT) - e^{-iK_n} e^{-iV_\mathrm{vV}^n NT} e^{iK_n} || &\leq& || U(NT) - e^{-iV_\mathrm{FM}^n NT} || + || e^{-iK_n} e^{-iV_\mathrm{vV}^n NT} e^{iK_n} - e^{-iV_\mathrm{FM}^n NT} || \nonumber \\
&=& e^{-O(\tilde{\omega})}+O(\tilde{\omega}^{-n-1}) \label{error2}.
\end{eqnarray}
Since the error $|| U(NT) - e^{-iK_n} e^{-iV_\mathrm{vV}^n NT} e^{iK_n} ||$ determines how long the effective Hamiltonian $D_n$ is valid, we provide the detailed analysis of the upper bound in the section \textbf{\ref{Sec1-3}}.

Finally, what should be done is to prove that any truncated van Vleck effective Hamiltonian commutes with the $\mathbb{Z}_N$ symmetry operation $X$. In the interaction picture, the Fourier component $V_m$ is calculated as follows,
\begin{eqnarray}
V_m &=& \frac{1}{NT} \int_0^{NT}  V_\mathrm{int}(t) e^{im\omega t/N} dt \nonumber \\
&=& \frac{1}{NT} \sum_{l=0}^{N-1} \int_0^T U_0^\dagger(t+lT) V(t) U_0(t+lT) e^{im \omega (t+lT)/N}  dt \nonumber \\
&=& \frac{1}{NT} \sum_{l=0}^{N-1} X^{-l}  \int_0^T V_\mathrm{int}(t)  e^{im \omega (t+lT)/N}dt \, X^l.
\end{eqnarray}
Thus, by using the condition $X^N=1$, the action of $X$ on the Fourier component is
\begin{eqnarray}
XV_mX^{-1} &=& \frac{1}{NT} \sum_{l=0}^{N-1} X^{-l+1}  \int_0^T V_\mathrm{int}(t)  e^{im \omega (t+lT)/N}dt \, X^{l-1} \nonumber \\
&=& e^{i2\pi m/N} V_m. \label{PhaseV}
\end{eqnarray}
When we focus on the $n$-th order terms of the van Vleck expansion, each term is a product of $n$ Fourier components, that is, it is composed of,
\begin{equation}
V_{m_1}V_{m_2}\hdots V_{m_n}.
\end{equation}
From Eq. (\ref{PhaseV}), this $n$-th order component transforms under $X$ as follows,
\begin{equation}
XV_{m_1}V_{m_2}\hdots V_{m_n}X^{-1} = \exp \left( \frac{2\pi i}{N}\sum_{i=1}^n m_i \right) V_{m_1}V_{m_2}\hdots V_{m_n}. \label{PhaseVV}
\end{equation}

\begin{figure}[t]
\begin{center}
    \includegraphics[height=4cm, width=18cm, clip]{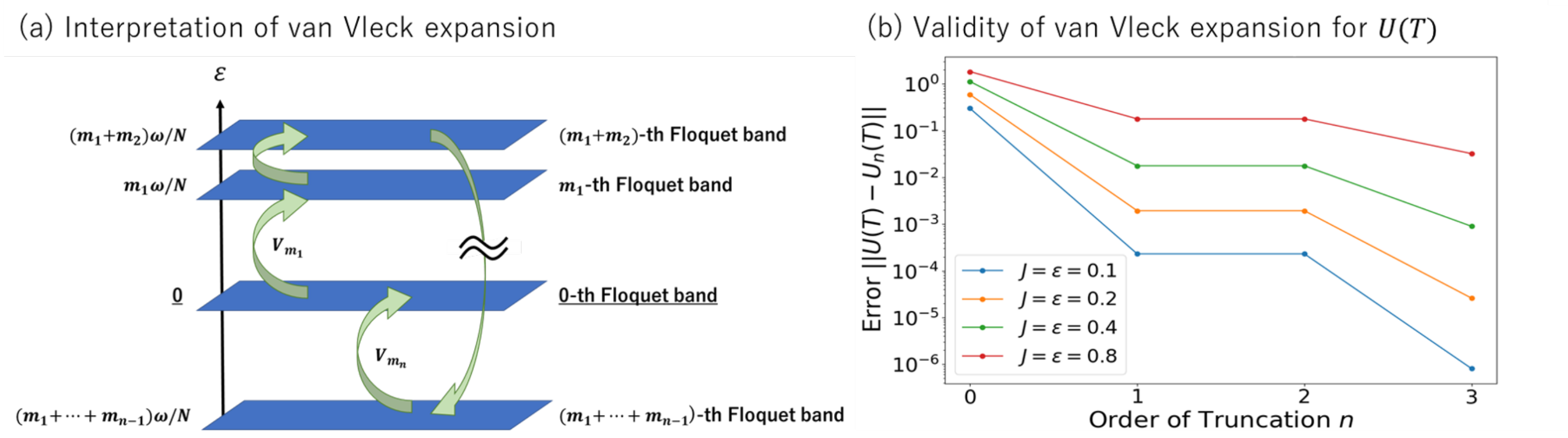}
    \caption{(a) Physical interpretation of each $n$-th order term of the van Vleck expansion. Each Fourier component $V_m$ gives rise to a jump to the $m_i$-th upper Floquet band, and each $n$-th order term represents a process where a state starts and ends at the 0-th Floquet band with $n$ times jumps. (b) The deviation $||U(T)-U_n(T)||$ for the model designated by Eqs. (\ref{Model})---(\ref{ModelVt-S}) for the system size $L=4$. As the truncation order $n$ or the renormalized frequency $\tilde{\omega}=1/\lambda NT$ increases, the error caused by the truncation becomes smaller.}
    \label{InterpretValidity}
  \end{center}
\end{figure}

We note here that, in the van Vleck effective Hamiltonian, which is derived by the van Vleck degenerate method \cite{Eckardt2015-S}, each $n$-th order term $V_{m_1}V_{m_2}\hdots V_{m_n}$ originates from at the virtual process, in which the state starts from the Floquet band whose energy is almost zero, jumps to the $m_i$ higher Floquet bands, and finally comes back to the original Floquet band after the $n$ times jumps (See Fig. \ref{InterpretValidity} (a)). Thus, concerning the indices of each $n$-th order term $V_{m_1}V_{m_2}\hdots V_{m_n}$,
\begin{equation}
\sum_{i=1}^n m_i = 0
\end{equation}
is satisfied. Therefore, from Eq. (\ref{PhaseVV}), $n$-th order terms in the van Vleck expansion commute with $X$ for arbitrary $n$. Since the effective Hamiltonian $V_\mathrm{vV}^n$ is obtained by truncating at a finite $n$-th order, the commutation relation
\begin{equation}
[V_\mathrm{vV}^n,X]=0
\end{equation}
is strictly satisfied. Then by regarding the van Vleck effective Hamiltonian $V_\mathrm{vV}^n$ as $D_n$ in the theorem, the theorem has been proved. \wsq

\subsection{Rigorous bound of the error and the lifetime}\label{Sec1-3}
In this section, we discuss the upper bound of the error $|| U(NT) - e^{-iK_n} e^{-iV_\mathrm{vV}^n NT} e^{iK_n} ||$ and thereby the lifetime of the prethermal regime, which corresponds to the time scale when the description by the truncated effective Hamiltonian $D_n$ becomes invalid. From Eq. (\ref{error2}),
\begin{equation}
|| U(NT) - e^{-iK_n} e^{-iV_\mathrm{vV}^n NT} e^{iK_n} || \leq || U(NT) - e^{-iV_\mathrm{FM}^n NT} || + || e^{-iK_n} e^{-iV_\mathrm{vV}^n NT} e^{iK_n} - e^{-iV_\mathrm{FM}^n NT} ||
\end{equation}
is satisfied under the assumptions (i)---(iii) in the theorem. When we regard the system as a $NT$-periodic Floquet system driven by $V_\mathrm{int}(t)$, the theorem by Kuwahara et al. \cite{Kuwahara2016-S} is applicable to the system. Thus, the upper bound of the first term is obtained as follows,
\begin{eqnarray}
|| U(NT) - e^{-iV_\mathrm{FM}^n NT} || &\leq& \frac{3}{k} L (\lambda NT) 2^{-\nu} + \frac{2(n-1)!}{n^2 k} L (\lambda NT)^{n+1} \nonumber \\
&\leq& \frac{3L}{2k} \cdot 2^{-1/16\lambda NT} + \frac{2L}{k} (n\lambda NT)^{n+1}
\end{eqnarray}
The upper bound of the second term $|| e^{-iK_n} e^{-iV_\mathrm{vV}^n NT} e^{iK_n} - e^{-iV_\mathrm{FM}^n NT} ||$ can be derived by the relationship between the truncated Floquet Magnus expansion and the truncated van Vleck expansion, which we discuss in the sections \textbf{\ref{SecFMvVRel}} and \textbf{\ref{SecKickExtensive}}. As a result, the second error term is bounded from above as follows (See \textbf{Lemma. 2} and Eq. (\ref{Sec32Result})):
\begin{equation}
||e^{-iK_n} e^{-iV_\mathrm{vV}^n NT} e^{iK_n} - e^{-iV_\mathrm{FM}^n NT}|| \leq
\frac{16L}{3k} (4n\lambda NT)^{n+1}
\left\{
\frac{j_{\mathrm{max},n}-(j_{\mathrm{max},n})^{n}}{1-j_{\mathrm{max},n}}+\frac{(j_{\mathrm{max},n})^{n}}{1-4j_{\mathrm{max},n}n\lambda NT}
\right\},
\end{equation}
if the period $T$ is small enough so that $4j_{\mathrm{max},n}n\lambda NT < 1$ is satisfied.
Here, $j_{\mathrm{max},n}$ is the renormalized extensiveness of $\{ K^{(i)} \}_{i=1}^n$, which is defined by Eqs. (\ref{jnDef}) and (\ref{JmaxDef}), and for example, they are bounded as follows (See the section \textbf{\ref{SecKickExtensive}}):
\begin{equation}\label{jmaxEg}
j_{\mathrm{max},1} \leq \frac{1}{2}, \quad j_{\mathrm{max},2} \leq \frac{1}{2}, \quad j_{\mathrm{max},3} \leq \frac{1}{2}.
\end{equation}
Therefore, we can obtain the upper bound for the error $|| U(NT) - e^{-iV_\mathrm{FM}^n NT} ||$ as
\begin{eqnarray}
|| U(NT) &-& e^{-iK_n} e^{-iV_\mathrm{vV}^n NT} e^{iK_n}  || \nonumber \\ 
&\leq& \frac{3L}{2k} \cdot 2^{-1/16\lambda NT} 
+ \frac{16L}{3k} (4n\lambda NT)^{n+1}
\left\{
\frac{j_{\mathrm{max},n}-(j_{\mathrm{max},n})^{n}}{1-j_{\mathrm{max},n}}+\frac{(j_{\mathrm{max},n})^{n}}{1-4j_{\mathrm{max},n}n\lambda NT}
+ \frac{3}{2^{2n+5}}
\right\}. \label{BoundResult}
\end{eqnarray}

Let us evaluate the right hand side of Eq. (\ref{BoundResult}) for simple cases. Based on Eq. (\ref{jmaxEg}), we consider the cases when the truncation order $n$ is small, and assume that $j_{\mathrm{max},1} \leq \frac{1}{2}$ (in fact, we can also discuss the cases when $j_{\mathrm{max},n}<1$ in the same way). The assumption that the truncation order $n$ is small is sensible for the analysis. Then, as long as $j_{\mathrm{max},1} \leq \frac{1}{2}$ is satisfied, we can provide a simple upper bound of the error from Eq. (\ref{BoundResult}) as follows:
\begin{equation}\label{SimpleBound}
|| U(NT) - e^{-iK_n} e^{-iV_\mathrm{vV}^n NT} e^{iK_n}  || \leq \frac{3L}{2k} \cdot 2^{-1/16\lambda NT} 
+ \frac{32L}{3k} (4n\lambda NT)^{n+1}.
\end{equation}
Since the function $(4n\lambda NT)^{n+1}$ is small compared to $1$ when $4n\lambda NT \lesssim 1$ or equivalently $n<O(1/\lambda NT)$ is satisfied, the right hand side is small enough as long as $\lambda NT$ and $n$ are small compared to $1$ and $\nu=O(1/\lambda NT)$, respectively. Therefore, the inequality Eq. (\ref{SimpleBound}) ensures the validity of the approximated time evolution operator $e^{-iK_n} e^{-iV_\mathrm{vV}^n NT} e^{iK_n}$.

The inequality Eq. (\ref{SimpleBound}) can also provide the rigorous lower bound for the lifetime of the prethermal regime. Let us define the lifetime $\tau_\ast^{(n)}$ by the time scale which validates the description by the $n$-th order truncated effective Hamiltonian $D_n$. Then, the lifetime is given by
\begin{equation}\label{DefLifetime}
|| U(NT) - e^{-iK_n} e^{-iV_\mathrm{vV}^n NT} e^{iK_n} || \cdot \frac{\tau_\ast^{(n)}}{NT} \sim 1,
\end{equation}
and we can rigorously evaluate the lower bound by using Eq. (\ref{SimpleBound}) as follows:
\begin{equation}\label{LifeBound}
\tau_\ast^{(n)} \gtrsim \frac{kNT}{L} \min \left\{ \frac{2}{3} \cdot 2^{1/16\lambda NT}, \frac{3}{32} (4n\lambda NT)^{-n-1} \right\}.
\end{equation}
Thus, we can describe the effective dynamics well by the truncated effective Hamiltonian $D_n$ at least in the time scale given by this bound.

Finally, we would like to note that the bounds for the error and the lifetime given by Eqs. (\ref{SimpleBound}) and (\ref{LifeBound}) are rigorous ones to ensure the validity of the effective Hamiltonian for general systems. Thus, it is possible that the error becomes larger than the bound given by Eq. (\ref{SimpleBound}), or equivalently, the lifetime becomes smaller than the bound given by Eq. (\ref{LifeBound}). In fact, in the simple example described in the main text, the error $|| U(NT) - e^{-iK_n} e^{-iV_\mathrm{vV}^n NT} e^{iK_n}  ||$ is much smaller than expected by Eq. (\ref{SimpleBound}) (See Fig. 1. (a) in the main text), and hence the lifetime of the prethermal discrete time crystalline order is much longer than expected by Eq. (\ref{LifeBound}) (See Fig. 2. (c) in the main text). More detailed evaluation of the error and the lifetime of the prethermal regime is left for future work.

\begin{figure}[t]
\begin{center}
    \includegraphics[height=5.5cm, width=18cm, clip]{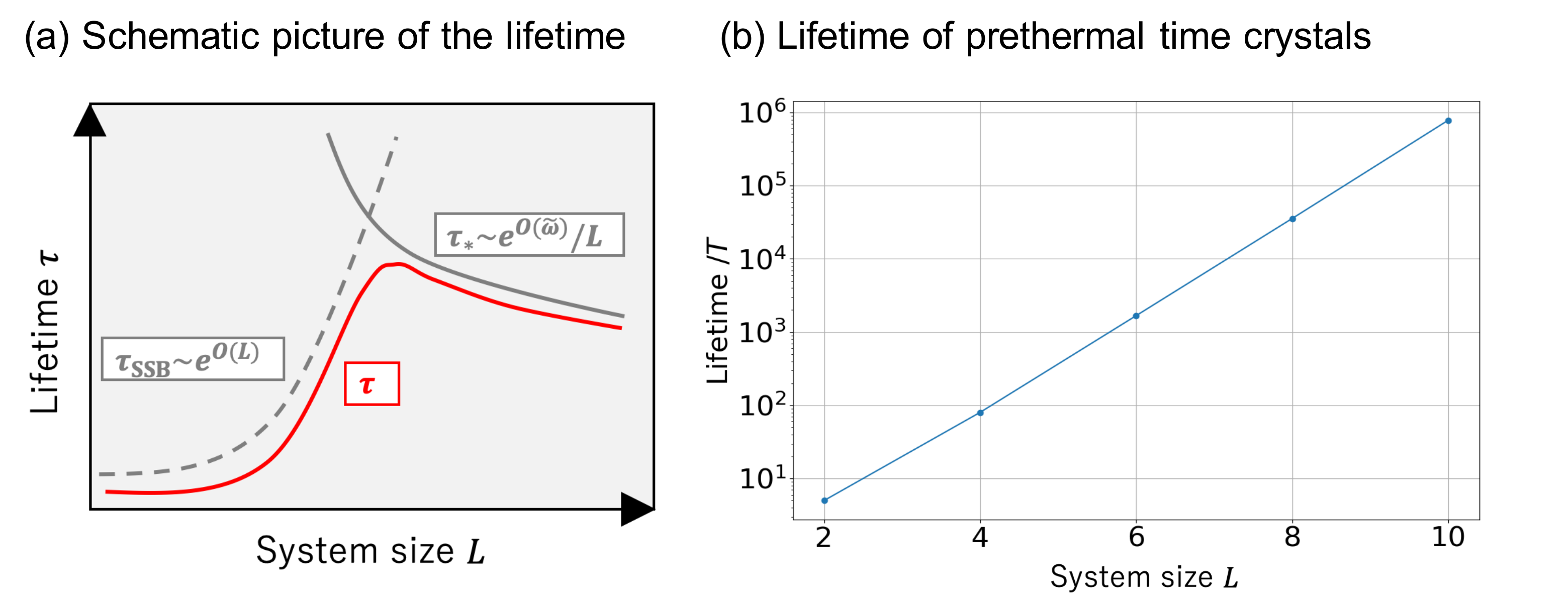}
    \caption{(a) Schematic picture of the lifetime of the prethermal time crystals. The smaller one of the two time  scales, $\tau_\mathrm{SSB}$ (gray dashed line) and $\tau_\ast$ (gray solid line) determines the lifetime of prethermal time crystals, $\tau$ (red solid line). (b) Numerical results for the lifetime of prethermal time crystals, which are described by Eqs. (1), (9), and (10) in the main text. The parameters $J$ (interaction) and $\varepsilon$ (transverse field) are chosen by $J=\varepsilon=0.9$ with the period $T=1$.}
    \label{LifetimeSize}
  \end{center}
\end{figure}

\subsection{Lifetime of prethermal discrete time crystals}

Here, we would like to briefly describe how the system size $L$ has an impact on the lifetime of prethermal time crystals, denoted by $\tau$.

As discussed in the main text, time crystalline orders in prethermal regimes are accompanied by spontaneous  breaking of the emergent $\mathbb{Z}_N$ symmetry $X$ under the effective static Hamiltonian $D_n$. Thus, when we consider the finite system size $L$, there are two important time scale---one is the lifetime of the prethermal regime $\tau_\ast$, and the other is the lifetime of the ordered phases brought by the spontaneous breaking of the symmetry $X$ under the effective static Hamiltonian, which we denote by $\tau_\mathrm{SSB}$. 

Assume that $\tau_\ast$ is smaller than $\tau_\mathrm{SSB}$. Then,  the ordered phases under the effective Hamiltonian are maintained as long as the system lies in the prethermal regime, and hence $\tau_\ast$ determines the lifetime of prethermal time crystals. On the contrary, when $\tau_\ast$ is larger than $\tau_\mathrm{SSB}$, $\tau_\mathrm{SSB}$ determines the lifetime of prethermal time crystals. Therefore, the lifetime of prethermal time crystals, $\tau$, is expected to be given as follows:

\begin{equation}
\tau \sim \min \left\{ \tau_\ast , \tau_\mathrm{SSB} \right\}.
\end{equation}

In general, $\tau_\mathrm{SSB}$, the lifetime of the ordered phases in an effective static system, is expected to exponentially grow with the system  size $L$. On the other hand, as you can see in Eq. (\ref{LifeBound}), the lifetime of the prethermal regime, $\tau_\ast$, decreases in proportion to $L^{-1}$, although the lifetime itself is exponentially long with the renormalized frequency $\tilde{\omega}=1/\lambda NT$. Thus, the lifetime of prethermal time crystals is approximately influenced by the system size as follows.

\begin{equation}
\tau \sim \min \left\{ \frac{e^{O(\tilde{\omega})}}{L}, e^{O(L)} \right\}.
\end{equation}

We describe the schematic picture of this size-dependence of the lifetime $\tau$, as shown in Fig. \ref{LifetimeSize} (a). We also numerically calculate the size-dependence of prethermal time crystals with the example described by Eq. (1), (9), and (10) in the main text (See Fig. \ref{LifetimeSize} (b)).  For the small system size $L$, which is accessible numerically by the exact diagonalization or experimentally in trapped ions \cite{Zhang2017-S}, the lifetime of prethermal time crystals exponentially increases with the system size $L$. This represents that, in microscopic or mesoscopic systems, the lifetime of the ordered phases brought by the spontaneous breaking of the symmetry $X$ is dominant, and hence it is difficult to distinguish prethermal time crystals from genuine time crystals by the finite size effect in microscopic or mesoscopic systems.


\section{Relation to the theorem by Else et al.}
In the main text, we introduce the theorem which provides the effective Hamiltonian and the existence of the new emergent symmetry for Floquet systems described by Eq. (\ref{Hamiltonian-S}).  Else et al. also consider similar Floquet systems described by Eqs. (\ref{Hamiltonian-S}) and (\ref{SymmetryOp-S}), and they also obtain approximate formulas for the time evolution operator and the existence of a new emergent symmetry. In this section, we would like to discuss the relationship between our theorem and the theorem given by Else et al. \cite{Else2017-S}.

First, let us briefly introduce what the theorem by Else et al. means. We consider Floquet systems which satisfy the conditions (i) and (ii) in the theorem, corresponding to Eqs. (\ref{Hamiltonian-S}) and (\ref{SymmetryOp-S}). In addition, assume that the local energy scale of $V(t)$, defined by $\lambda$, is finite. Though the definition of the norm for operators is a little different from ours, physical interpretations of the theorems are essentially the same for both cases. When the energy scale of $V(t)$ is much smaller than the frequency $\omega$, which means $\lambda NT \ll 1$, their theorem dictates that the Floquet operator for the systems is approximated as
\begin{eqnarray}
U(T) &=& \mathcal{U} X e^{-iDT} \mathcal{U}^\dagger +O\left( e^{O(\tilde{\omega}/(\log \tilde{\omega})^3)} \right), \label{U_T}\\
D &=& V_0 + \lambda O\left( (\lambda NT)^2 \right), \label{D_EBN}\\
\mathcal{U} &=& 1+O(\lambda NT), \label{U_EBN}
\end{eqnarray}
 where $\mathcal{U}$ is unitary and the effective Hamiltonian $D$ acquires the new emergent symmetry generated by $X$, that is,
\begin{equation}
[D,X]=0.
\end{equation}

The theorem by Else et al. provides the time evolution operator for long time scale $t<\exp(O(\tilde{\omega}/(\log \tilde{\omega})^3))$, and the existence of the emergent symmetry generated by $X$, and thereby it ensures the possibility of discrete time crystalline order in the prethermal regime as well as our theorem. Here, we would like to discuss differences between the theorems. Note that one of these theorems does not include the other.

The first one is that our theorem provides the time evolution operator $U(NT)$, whereas Else's theorem provides the Floquet operator $U(T)$. Therefore, though Else's theorem tells us the stroboscopic dynamics every period $T$, our theorem predicts only the coarse-grained stroboscopic dynamics every $N$ periods $NT$. However, our theorem can ensure the possibility of discrete time crystal orders in the prethermal regime as well as Else's theorem. The second important difference is that our formalism provides the effective Hamiltonian and the unitary transformation which describe the effective dynamics with the emergent symmetry including higher order correction terms, though Else's theorem gives only the lowest order as Eqs. (\ref{D_EBN}) and (\ref{U_EBN}). 

As we have described in the main text, the second difference, which we would like to emphasize the most, enables the detailed analysis of discrete time crystalline orders in the prethermal regime and the simultaneous control of phases and symmetries by a periodic drive. Concerning the analysis of prethermal DTCs, the high frequency expansion for the effective Hamiltonian and the unitary transformation are of use for the determination of the condition for the phase transition and the effective inverse temperature. In particular, we can also prepare the effective zero temperature state $e^{iK_n}\ket{\psi_0}$, where $\ket{\psi_0}$ is a symmetry-broken ground state of the effective Hamiltonian $D_n$. Thus, our formalism makes it possible to realize prethermal DTCs in one-dimensional systems, as we have described in the main text. As well, the higher order terms realize an essentially new effective Hamiltonian, and hence we can perform the simultaneous control of phases and symmetries as an application of our formalism to Floquet engineering.

Concerning these two theorems, the ultimate goal is to obtain the effective Hamiltonian and the unitary transformation, given by the high frequency expansion, for the Floquet operator $U(T)$. Here, we examine such a theory from our theorem and Else's theorem, and numerically confirm it with a simple example. The theorem by Else et al., which gives an approximate formula for $U(T)$, can also provide an approximate formula for $U(NT)$ from Eq. (\ref{U_T}) as follows,
\begin{equation}
U(NT)=U(T)^N \sim \mathcal{U} e^{-iDNT} \mathcal{U}^\dagger, \qquad [D,X]=0.
\end{equation}
Since this equation should be consistent with Eq. (\ref{U_NT}), which is obtained from our theorem, the relations
\begin{equation}\label{Assumption}
D \simeq D_n, \qquad \mathcal{U} \simeq e^{-iK_n}
\end{equation}
are expected. Here, we numerically examine these relations. The model is a one-dimensional Ising chain under a periodic pulse and a periodic transverse field, described as
\begin{eqnarray}
H(t) &=& H_0(t)+V(t) \label{Model}\\
H_0(t) &=& \frac{\pi}{2} \sum_i \sigma_i^x \sum_{n \in \mathbb{Z}} \delta(t-nT), \label{ModelH0-S}\\
V(t) &=& -J\sum_i \sigma_i^z \sigma_{i+1}^z +\varepsilon \sin \omega t \sum_i \sigma_i^x. \label{ModelVt-S}
\end{eqnarray}
Then, $\lambda$, defined by the local energy scale of $V(t)$, is given by $\lambda\sim \max(J,\varepsilon)$. If the assumption of Eq. (\ref{Assumption}) is correct, $U(T)$ would be approximately equal to 
\begin{equation}
U_n(T) \equiv e^{-iK_n} X e^{-iD_nT} e^{iK_n}
\end{equation}
from Eq. (\ref{U_T}). Thus, we evaluate the deviation defined by the operator norm $||U(T)-U_n(T)||$, shown in Fig. \ref{InterpretValidity} (b). From this figure, the deviation $||U(T)-U_n(T)||$ roughly shows exponential decay with the truncation order $n$, or decreases as a power of $\lambda NT$. This dependency is exactly the same as that of the deviation $U(NT)-U_n(NT)$ described in Fig. 1 (a) in the main text. Therefore, it implies that the validity of the van Vleck expansion for $U(T)$, 
\begin{equation}
U(T) \simeq e^{-iK_n} X e^{-iD_nT} e^{iK_n},
\end{equation}
is confirmed at least in this simple example. An extension of the theory to general cases, which can give the effective Hamiltonian for the Floquet operator $U(T)$ including high order corrections, is left for future work. 

\section{High frequency expansion in Floquet systems}\label{SecExpansion}

\subsection{Explicit formulas for the high frequency expansions}
Here, we provide high frequency expansions--- the Floquet Magnus expansion and the van Vleck expansion, and their properties. We assume that the Hamiltonian of the system is $V_\mathrm{int}(t)$, whose period is $NT$. We denote the characteristic energy scale of the system by $\lambda$, which is defined by the condition (ii) in the theorem (Replace $V(t)$ by $V_\mathrm{int}(t)$). First of all, we define the $m$-th Fourier component $V_m$ by

\begin{equation}
V_m = \frac{1}{NT} \int^{NT}_0 V_\mathrm{int}(t) e^{im\omega t/N} dt,
\end{equation}
where the frequency $\omega$ is equal to $2\pi/T$.

First, the Floquet Magnus effective Hamiltonian $V_\mathrm{FM}$ is defined by the time evolution operator for one period as follows,

\begin{equation}\label{DefFloquetMagnus}
U(NT)\equiv e^{-iV_\mathrm{FM}NT}, \quad \mathrm{where} \quad U(NT) = \mathcal{T} \mathrm{exp} \left( -i \int^{NT}_0 V_\mathrm{int}(t')dt' \right).
\end{equation}
Then, the Floquet Magnus expansion is defined by the perturbation expansion of Eq. (\ref{DefFloquetMagnus}) for the small parameter $\lambda NT$ at high frequency. Each $n$-th order term $V_\mathrm{FM}^{(n)}$ is obtained as follows,

\begin{eqnarray}
V_\mathrm{FM}^{(0)} &=& 0, \label{FloquetMagnus0} \\
V_\mathrm{FM}^{(1)} &=& V_0, \label{FloquetMagnus1} \\
V_\mathrm{FM}^{(2)} &=& N\sum_{m\neq0}\frac{[V_{-m},V_m]}{2m\omega} + N\sum_{m \neq 0} \frac{[V_m,V_0]}{m\omega}, \label{FloquetMagnus2} \\
V_\mathrm{FM}^{(3)} &=& N^2\sum_{m\neq0}\frac{[[V_{-m},V_0],V_m]}{2m^2\omega^2}+N^2\sum_{m\neq0}\sum_{n\neq0,m} \frac{[[V_{-m},V_{m-n}],V_n]}{3mn\omega^2}-N^2 \sum_{m\neq 0}\frac{[[V_m,V_0],V_0]}{m^2\omega^2} \nonumber \\
&\quad& \quad + N^2 \sum_{m,n\neq 0} \frac{-[[V_m,V_n],V_{-n}]+[[V_n,V_{-n}],V_m]}{3mn\omega^2} - N^2\sum_{m\neq0}\sum_{n\neq0,m} \frac{[[V_{n},V_{m-n}],V_0]}{2mn\omega^2} \nonumber \\
&\quad& \quad \quad + N^2 \sum_{m,n\neq 0} \frac{[[V_m,V_n],V_0]-[[V_m,V_0],V_n]}{2mn\omega^2}. \label{FloquetMagnus3} \\
&\vdots& \nonumber
\end{eqnarray}
Here, we determine the order of the expansion $n$ so that $V_\mathrm{FM}^{(n)}NT=O((\lambda NT)^n)$ is satisfied. We also define the truncated Floquet Magnus effective Hamiltonian by
\begin{equation}
V_\mathrm{FM}^n = \sum_{i=0}^n V_\mathrm{FM}^{(i)}
\end{equation}
for $n \in \mathbb{N}$. Though the Floquet Magnus expansion does not necessarily converge in generic many-body periodically driven systems, we can analyze the system by the truncated Floquet Magnus effective Hamiltonian $V_\mathrm{FM}^\nu$ in high frequency regime $\lambda NT \ll 1$ thanks to the theorem by Kuwahara et al.\cite{Kuwahara2016-S,Mori2016-S}.

However, in general, the Floquet Magnus effective Hamiltonian $V_\mathrm{FM}$ depends on the choice of the Floquet gauge (the time origin), that is, the effective Hamiltonian $V_\mathrm{FM}$ varies if we change the time origin from $t=0$. The effective Hamiltonian transformed by a unitary operation so that it becomes independent of the Floquet gauge is the van Vleck effective Hamiltonian $V_\mathrm{vV}$, which is defined as

\begin{equation}\label{DefEffective}
V_\mathrm{FM} = e^{-iK} V_\mathrm{vV} e^{iK}.
\end{equation}
Here, The hermitian operator $K$, which gives the unitary transformation between the van Vleck effective Hamiltonian and the Floquet Magnus effective Hamiltonian, is called a kick operator. By this unitary transformation, only the kick operator depends on the Floquet gauge. The expansions of the van Vleck effective Hamiltonian and the kick operator can be obtained by the van Vleck degenerate method \cite{Eckardt2015-S,Bukov2015-S,Mikami2016-S}, which results in

\begin{eqnarray}
V_\mathrm{vV}^{(0)} &=& 0, \label{vanVleck0} \\
V_\mathrm{vV}^{(1)} &=& V_0, \label{vanVleck1} \\
V_\mathrm{vV}^{(2)} &=& N\sum_{m\neq0}\frac{[V_{-m},V_m]}{2m\omega}, \label{vanVleck2} \\
V_\mathrm{vV}^{(3)} &=& N^2\sum_{m\neq0}\frac{[[V_{-m},V_0],V_m]}{2m^2\omega^2}+N^2\sum_{m\neq0}\sum_{n\neq0,m} \frac{[[V_{-m},V_{m-n}],V_n]}{3mn\omega^2} \label{vanVleck3}, \\
&\vdots&, \nonumber \\
K^{(0)} &=& 0 \label{kick0}, \\
iK^{(1)} &=& -N\sum_{m\neq0}\frac{V_m}{m\omega} \label{kick1}, \\
iK^{(2)} &=& N^2\sum_{m\neq0}\sum_{n\neq0,m} \frac{[V_m,V_{m-n}]}{2mn\omega^2} + N^2\sum_{m\neq0} \frac{[V_m,V_0]}{m^2\omega^2}, \label{kick2}\\
iK^{(3)} &=& -N^3 \sum_{m \neq 0} \frac{[[V_m,V_0],V_0]}{m^3\omega^3}+N^3 \sum_{m \neq 0}\sum_{n \neq 0}\frac{[V_m,[V_{-n},V_n]]}{4m^2n\omega^3} \nonumber \\
&\quad& \quad - N^3 \sum_{m \neq 0}\sum_{n \neq 0,m}\left( \frac{[[V_n,V_0],V_{m-n}]}{2mn^2\omega^3}+\frac{[[V_n,V_{m-n}],V_0]}{2m^2n\omega^3} \right) \nonumber \\
&\quad& \quad \quad - N^3 \sum_{m \neq 0}\sum_{n \neq 0}\sum_{l \neq 0,m,n} \frac{[V_n,[V_{l-n},V_{m-l}]]}{4mnl\omega^3} - N^3 \sum_{m \neq 0}\sum_{n \neq 0}\sum_{l \neq 0,m-n} \frac{[V_n,[V_{l},V_{m-n-l}]]}{12mnl\omega^3}. \label{kick3} \\
&\vdots& \nonumber
\end{eqnarray}
We also define the truncated van Vleck effective Hamiltonian $V_\mathrm{vV}^\nu$ and the truncated kick operator $K_\nu$ by
\begin{equation}
V_\mathrm{vV}^n = \sum_{i=0}^n V_\mathrm{vV}^{(i)}, \quad K_n = \sum_{i=0}^n K^{(i)}.
\end{equation}

\subsection{Relationship between the two different schemes of the truncated high frequency expansions}\label{SecFMvVRel}
Here, we would like to discuss the relationship between the truncated Floquet Magnus expansion and the truncated van Vleck expansion, which determines the validity of the effective Hamiltonian in the main theorem as described in the section \ref{Derivation}. In other words, we evaluate the error $||e^{-iK_n} e^{-iV_\mathrm{vV}^n NT} e^{iK_n} - e^{-iV_\mathrm{FM}^n NT}||$ in this section. First, because of the following lemma, we can evaluate the error by the upper bound of $||e^{-iK_n}V_\mathrm{vV}^n e^{iK_n}-V_\mathrm{FM}^n||$.

\begin{lemma}
For any bounded hermitian operators $A$ and $B$,
\begin{equation}\label{Lemma1Result}
|| e^{iA}-e^{iB} || \leq 2 ||A-B||
\end{equation}
is satisfied. 
\end{lemma}
\begin{proof*}
For a bounded hermitian operator $C$ and $t \in \mathbb{R}$, we calculate the upper bound of $||F(t)-F(0)||$, where $F(t)$ is a unitary operator defined by $F(t)=\exp \{i(B+Ct)\}$. By denoting $F^{(n)}(t)=d^n F(t)/dt^n$,
\begin{equation}
|| F(t)-F(0) || \leq \left| \left| \sum_{n=1}^{\infty} \frac{1}{n!} F^{(n)}(0) t^n \right| \right| \leq \sum_{n=1}^{\infty} \frac{|t|^n}{n!} || F^{(n)}(0) ||
\end{equation}
is obtained. By using the formula
\begin{equation}
\frac{d}{dt}e^{X(t)} = \int_{0}^{1} d\alpha e^{\alpha X(t)} \left( \frac{d}{dt} X(t) \right) e^{(1-\alpha)X(t)},
\end{equation}
we can evaluate $||F^{(1)}(t)||$ as follows:
\begin{equation}
||F^{(1)}(t)|| \leq \left| \left| \int_0^1d\alpha e^{i\alpha(B+Ct)} iC e^{i(1-\alpha)(B+Ct)} \right| \right| \leq ||C||.
\end{equation}
In a similar way,
\begin{eqnarray}
||F^{(2)}(t)|| &\leq& \iint_0^1 \alpha \left| \left| e^{i\beta\alpha(B+Ct)} i C e^{i(1-\beta)\alpha(B+Ct)} iC e^{i(1-\alpha)(B+Ct)} \right| \right| d\alpha d\beta \nonumber \\
&\quad& \qquad + \iint_0^1 (1-\alpha) \left| \left| e^{i\alpha(B+Ct)} iC e^{i\beta(1-\alpha)(B+Ct)} iC e^{i(1-\beta)(1-\alpha)(B+Ct)} \right| \right| d\alpha d\beta \\
&\leq& ||C||^2
\end{eqnarray}
is obtained. By repeating this calculation, $||F^{(n)}(t)||\leq ||C||^n$ is satisfied. Therefore, if we assume $||C||\cdot |t| \leq 1$, we get
\begin{equation}
|| F(t)-F(0) || \leq \sum_{n=1}^{\infty} \frac{1}{n!} ||C||^n \cdot |t|^n \leq ||C||\cdot |t| \sum_{n=1}^\infty \frac{1}{n!} \leq 2 || C || \cdot |t|.
\end{equation}
This inequality is satisfied also when $||C||\cdot |t| \geq 1$, because of
\begin{equation}
|| F(t)-F(0) ||=||e^{i(B+Ct)}-e^{iB}|| \leq ||e^{i(B+Ct)}||+||e^{iB}|| \leq 2 \leq 2 ||C|| \cdot |t|.
\end{equation}
Finally, if we substitute $C=A-B$ and $t=1$, we obtain the inequality Eq. (\ref{Lemma1Result}). \wsq
\end{proof*}

If we substitute $A=-e^{-iK_n} V_\mathrm{vV}^n NT e^{iK_n}$ and $B=-V_\mathrm{FM}^n NT$, we can obtain
\begin{equation}
||e^{-iK_n} e^{-iV_\mathrm{vV}^n NT} e^{iK_n} - e^{-iV_\mathrm{FM}^n NT}|| \leq 2NT ||e^{-iK_n}V_\mathrm{vV}^n e^{iK_n}-V_\mathrm{FM}^n||.
\end{equation}
Thus, it is sufficient to evaluate $||e^{-iK_n}V_\mathrm{vV}^n e^{iK_n}-V_\mathrm{FM}^n||$ instead of the error $||e^{-iK_n} e^{-iV_\mathrm{vV}^n NT} e^{iK_n} - e^{-iV_\mathrm{FM}^n NT}||$. To this end, we define the locality and the extensiveness of operators and describe some properties. First, if an operator $A$ contains at most $k$-body interactions, $A$ is called $k$-local, or $k$ is called the locality of $A$. In addition, an operator $A$ is $J$-extensive if $A$ satisfies
\begin{equation}
\sum_{X: i \in X} ||a_X|| \leq J, \quad ^\forall i,
\end{equation}
where $a_X$ represents the terms in $A$ which nontrivially act on the domain $X$. If $A$ is $J$-extensive, we can obtain 
\begin{eqnarray}
||A|| &\leq& \sum_X || a_X || \nonumber \\
&\leq& \sum_i \sum_{X: i \in X} ||a_X || \nonumber\\
&\leq& LJ, \label{ALJ}
\end{eqnarray}
where $L$ is the system size. When a series of operators $A_i$ are $k_i$-local and $J_i$-extensive respectively, the extensiveness of the multi-commutator $J_{\{A_n,A_{n-1},\hdots,A_1\}}$ is bounded from above as
\begin{eqnarray}
\{A_n,A_{n-1},\hdots,A_1\} &\equiv& [A_n,[A_{n-1},\hdots,[A_2,A_1]]\hdots] \\
J_{\{A_n,A_{n-1},\hdots,A_1\}} &\leq& J_1 \prod_{i=2}^n (2J_i K_i), \label{ExtensiveRule}
\end{eqnarray} 
where $K_i = \sum_{m \leq i} k_m$ (See Lemma. 5 in Ref. \cite{Kuwahara2016-S}). With this property, we can obtain the following lemma, which gives the upper bound of $||e^{-iK_n}V_\mathrm{vV}^n e^{iK_n}-V_\mathrm{FM}^n||$.

\begin{lemma}
Consider a Floquet system which satisfies all of the conditions (i)---(iii) in the main theorem, and denote $\tilde{T} \equiv NT$. We also denote the extensiveness of the $n$-th order term of the kick operator $K^{(n)}$ by $J^{(n)}$. Then, for the truncation order $n$ smaller than $\nu=O(1/\lambda \tilde{T})$,
\begin{equation}\label{LemmaResult}
||e^{-iK_n}V_\mathrm{vV}^n e^{iK_n}-V_\mathrm{FM}^n|| \leq \frac{8L}{3k\tilde{T}} (4n\lambda \tilde{T})^{n+1}
\left\{
\frac{j_{\mathrm{max},n}-(j_{\mathrm{max},n})^{n}}{1-j_{\mathrm{max},n}}+\frac{(j_{\mathrm{max},n})^{n}}{1-4j_{\mathrm{max},n}n\lambda \tilde{T}}
\right\}
\end{equation}
is satisfied when the period $T$ is small enough to satisfy $4j_{\mathrm{max},n}n\lambda T < 1$, where the maximal value of the renormalized extensiveness $j_{\mathrm{max},n}$ is defined as follows:
\begin{eqnarray}
j^{(n)} &\equiv& \frac{J^{(n)}}{(\lambda \tilde{T})^{n-1}  J\tilde{T} n^n}, \label{jnDef} \\
j_{\mathrm{max},n} &\equiv& \max \{ j^{(l)} | \, 1 \leq l \leq n \}. \label{JmaxDef}
\end{eqnarray}
\end{lemma}

\begin{proof*}
By using the Baker-Campbell-Hausdorff formula, we obtain
\begin{eqnarray}
||e^{-iK_n}V_\mathrm{vV}^n e^{iK_n}-V_\mathrm{FM}^n|| &=& ||e^{iK_n}V_\mathrm{FM}^n e^{-iK_n}-V_\mathrm{vV}^n|| \nonumber \\
&=& \left| \left| \sum_{m=0}^\infty \frac{i^m}{m!} \sum_{\{l_i\}_{i=1}^m, 1 \leq l_i \leq n} \sum_{1 \leq l \leq n} \{ K^{(l_1)},K^{(l_2)},\hdots,K^{(l_m)},V_\mathrm{FM}^{(l)} \}- V_\mathrm{vV}^n \right| \right| \nonumber \\
&=& \left| \left| \left( \sum_{m=1}^{n-1} \sum_{M=n+1}^{(m+1)n} + \sum_{m=n}^\infty \sum_{M=m+1}^{(m+1)n} \right) \sum_{\substack{1\leq l_i,l \leq n \\ \sum_i l_i+l = M }} \frac{i^m}{m!} \{ K^{(l_1)},\hdots,K^{(l_m)},V_\mathrm{FM}^{(l)} \} \right| \right| \nonumber \\
&\leq& \left( \sum_{m=1}^{n-1} \sum_{M=n+1}^{(m+1)n} + \sum_{m=n}^\infty \sum_{M=m+1}^{(m+1)n} \right) \sum_{\substack{1\leq l_i,l \leq n \\ \sum_i l_i+l = M }} \frac{1}{m!}
\left| \left| \{ K^{(l_1)},\hdots,K^{(l_m)},V_\mathrm{FM}^{(l)} \} \right| \right| \label{Tochu1}
\end{eqnarray}
To derive the third equality, we use the fact that $e^{iK_n} V_\mathrm{FM}^n e^{-iK_n}$ is equivalent to $V_\mathrm{vV}^n$ up to the $n$-th order in $\lambda \tilde{T}$. When the Hamiltonian $V_\mathrm{int}(t)$ is $k$-local and $J$-extensive, the extensiveness of the $n$-th order Floquet Magnus expansion $J_{V_\mathrm{vV}^{(n)}}$ is bounded as follows \cite{Kuwahara2016-S},
\begin{equation}\label{FMextensive}
J_{V_\mathrm{FM}^{(n)}} \leq \frac{(\lambda \tilde{T})^{n-1}J}{n}(n-1)!, \quad \mathrm{where} \quad \lambda = 2kJ.
\end{equation}
Assume that the indices $l$ and $\{l_i\}$ satisfy $1 \leq l_i, l \leq n$ and $\sum_i l_i +l=M$, and note that the $n$-th order terms $K^{(n)}$ and $V_\mathrm{FM}^{(n)}$ are at most $nk$-local. Then, by using Eq. (\ref{ExtensiveRule}),
\begin{eqnarray}
\left| \left| \{ K^{(l_1)},\hdots,K^{(l_m)},V_\mathrm{FM}^{(l)} \} \right| \right| &\leq& L J_{\{ K^{(l_1)},\hdots,K^{(l_m)},V_\mathrm{FM}^{(l)} \}} \nonumber \\
&\leq& L J_{V_\mathrm{vV}^{(n)}} \prod_{i=1}^m (2J^{(l_i)}) \cdot (l+l_1)k \hdots (l+l_1+\hdots+l_m)k \nonumber \\
&\leq& LJ (\lambda \tilde{T})^{l-1} l! \prod_{i=1}^m \left\{ 2(\lambda \tilde{T})^{l_i-1} J\tilde{T} l_i^{l_i} j^{(l_i)} \right\} k^m (M-m+1)\hdots(M-1)M \nonumber \\
&=& LJ (\lambda \tilde{T})^{M-1} l! l_1^{l_1}l_2^{l_2}\hdots l_m^{l_m} (j_{\mathrm{max},n})^m m! \,_MC_m \nonumber \\
&\leq& LJ (\lambda \tilde{T})^{M-1} n^l n^{l_1} n^{l_2} \hdots n^{l_m} (j_{\mathrm{max},n})^m m! 2^M \nonumber \\
&=& 2nLJ (2n\lambda \tilde{T})^{M-1} (j_{\mathrm{max},n})^m m! \label{Tochu2}
\end{eqnarray}
is obtained. By using the assumption $n < \nu < 1/16\lambda \tilde{T}$ and $4j_{\mathrm{max},n}n\lambda \tilde{T}<1$, we end up with
\begin{eqnarray}
||e^{-iK_n}V_\mathrm{vV}^n e^{iK_n}-V_\mathrm{FM}^n|| &\leq& \left( \sum_{m=1}^{n-1} \sum_{M=n+1}^{(m+1)n} + \sum_{m=n}^\infty \sum_{M=m+1}^{(m+1)n} \right) \sum_{\substack{1\leq l_i,l \leq n \\ \sum_i l_i+l = M }} 2nLJ (2n\lambda \tilde{T})^{M-1} (j_{\mathrm{max},n})^m \nonumber \\
&\leq& \left( \sum_{m=1}^{n-1} \sum_{M=n+1}^{\infty} + \sum_{m=n}^\infty \sum_{M=m+1}^{\infty}  \right) 4nLJ (4n\lambda \tilde{T})^{M-1} (j_{\mathrm{max},n})^m \nonumber \\
&=& \frac{4nLJ(4n\lambda \tilde{T})^n}{1-4n\lambda \tilde{T}}
\left\{
\frac{j_{\mathrm{max},n}-(j_{\mathrm{max},n})^{n}}{1-j_{\mathrm{max},n}}+\frac{(j_{\mathrm{max},n})^{n}}{1-4j_{\mathrm{max},n}n\lambda \tilde{T}}
\right\} \label{Tochu3} \\
&\leq& \frac{8L}{3k\tilde{T}} (4n\lambda \tilde{T})^{n+1}
\left\{
\frac{j_{\mathrm{max},n}-(j_{\mathrm{max},n})^{n}}{1-j_{\mathrm{max},n}}+\frac{(j_{\mathrm{max},n})^{n}}{1-4j_{\mathrm{max},n}n\lambda \tilde{T}}
\right\}. \label{Tochu4}
\end{eqnarray}
Finally, we obtain the result of the lemma Eq. (\ref{LemmaResult}), and the proof has been completed. \wsq
\end{proof*}

By combining the results of \textbf{Lemma. 1} and \textbf{Lemma. 2}, we can evaluate the error $||e^{-iK_n} e^{-iV_\mathrm{vV}^n NT} e^{iK_n} - e^{-iV_\mathrm{FM}^n NT}||$ by
\begin{equation}
||e^{-iK_n} e^{-iV_\mathrm{vV}^n NT} e^{iK_n} - e^{-iV_\mathrm{FM}^n NT}|| \leq
\frac{16L}{3k} (4n\lambda \tilde{T})^{n+1}
\left\{
\frac{j_{\mathrm{max},n}-(j_{\mathrm{max},n})^{n}}{1-j_{\mathrm{max},n}}+\frac{(j_{\mathrm{max},n})^{n}}{1-4j_{\mathrm{max},n}n\lambda \tilde{T}}
\right\}, \label{Sec32Result}
\end{equation}
when the period $T$ is small enough to satisfy $4j_{\mathrm{max},n}n\lambda \tilde{T}<1$.

\subsection{Extensiveness of the kick operators}\label{SecKickExtensive}

Here, the extensiveness of the $n$-th order terms of the kick operators $K^{(n)}$, denoted by $J^{(n)}$, is discussed from low orders so that we can evaluate the error caused by the  transformation between the truncated Floquet Magnus expansion and the truncated van Vleck expansion, depicted as Eq. (\ref{Sec32Result}).

To this end, we use Eq. (\ref{ExtensiveRule}) and the following inequalities:
\begin{eqnarray}
\left| \sum_{m \neq 0} \frac{e^{im\omega t/N}}{m} \right| &\leq& \pi, \label{1/m} \\
\left| \sum_{m \neq 0} \frac{e^{im\omega t/N}}{m^s} \right| &\leq& \sum_{m \neq 0} \frac{1}{|m|^s} = 2 \zeta(s), \quad \mathrm{for} \quad s \in \mathbb{N}\backslash \{1\}. \label{1/m^s}
\end{eqnarray}
Here, $\zeta(s)$ is the zeta function. The first inequality is derived from the fact that
\begin{equation}
\sum_{m \neq 0} \frac{e^{im\omega t/N}}{m} =
\begin{cases}
0 & \mathrm{if} \quad \omega t/N=0 \mod 2\pi \\
i(\pi-\alpha) & \mathrm{if} \quad \omega t/N=\alpha \,\,\, (\in (0,2\pi)) \mod 2\pi.
\end{cases}
\end{equation}
From Eq. (\ref{kick1}),
\begin{equation}\label{Kick1New}
iK^{(1)} = -\frac{1}{\omega T} \int_0^{NT} \left( \sum_{m \neq 0} \frac{e^{im\omega t/N}}{m} \right) V_\mathrm{int}(t) dt
\end{equation}
is obtained, and hence the extensiveness of $K^{(1)}$ is bounded from above under the assumption that $V_\mathrm{int}(t)$ is $J$-extensive as Eq. (\ref{energy_bound}):
\begin{equation}\label{J1extensive}
J^{(1)} \leq \frac{1}{\omega T} \int_0^{NT} \left| \sum_{m \neq 0} \frac{e^{im\omega t/N}}{m} \right| J dt \leq \frac{1}{2} J\tilde{T}.
\end{equation}
Here, we have used the fact that, when a scalar function $f(t)$ satisfies $|f(t)|<F=\mathrm{Const.}$ and an operator $O(t)$ is $J_O$-extensive, the extensiveness, denoted by $J_{\bar{O}}$, of the operator $\bar{O} \equiv \int_0^T f(t) O(t)dt$ is bounded from above as $J_{\bar{O}}\leq FJ_OT$. This can be derived from
\begin{equation}
\sum_{X: i \in X} ||\bar{o}_X ||= \sum_{X: i \in X} \left| \left| \int_0^T f(t) o_X(t) dt \right| \right| \leq \int_0^T |f(t)| \sum_{X: i \in X} ||o_X(t)|| dt \leq FJ_OT,
\end{equation}
where we denote $O(t)=\sum_X o_X(t)$, and $\bar{O}=\sum_X \bar{o}_X$.
We can also discuss the second order terms $K^{(2)}$ given in Eq. (\ref{kick2}) as follows,
\begin{equation}\label{Kick2New}
iK^{(2)} =\frac{1}{2\omega^2 T^2} \iint_0^{NT} dt_1 dt_2 \left\{ \left( \sum_{m \neq 0} \frac{e^{im\omega (t_1+t_2)/N}}{m} \right) \left( \sum_{n \neq 0} \frac{e^{-in\omega t_2/N}}{n} \right)+ \left( \sum_{m \neq 0} \frac{e^{im\omega t_1/N}}{m^2} \right) \right\} [V_\mathrm{int}(t_1),V_\mathrm{int}(t_2)].
\end{equation}
This equation gives the upper bound of $J^{(2)}$, which can be described by
\begin{equation}\label{J2extensive}
J^{(2)} \leq \frac{1}{2\omega^2 T^2} \iint_0^{NT} dt_1 dt_2 \left( \pi^2 + 2\zeta(2) \right) J_{\{V_\mathrm{int}(t_1),V_\mathrm{int}(t_2)\}} \leq \frac{1}{3} (\lambda \tilde{T}) J \tilde{T}.
\end{equation}
In a similar way, by using $\zeta(3)=\pi^3/25.79\hdots<\pi^3/24$, we can also obtain the extensiveness of the third order as $J^{(3)} \leq 47(\lambda \tilde{T})^2 J\tilde{T}/96$. Finally, we can evaluate $j_{\mathrm{max},n}$, which is defined by Eqs. (\ref{jnDef}) and (\ref{JmaxDef}), as follows:
\begin{equation}\label{Jmax1123}
j_{\mathrm{max},1} \leq \frac{1}{2}, \quad j_{\mathrm{max},2} \leq \frac{1}{2}, \quad j_{\mathrm{max},3} \leq \frac{1}{2}.
\end{equation}

For a general integer $n$, the $n$-th order term of the kick operator $K^{(n)}$ can be decomposed into the inverse Flourier transformation of $1/m^s$ and the $n$-tuple commutator of $V_\mathrm{int}(t_i)$ like Eqs. (\ref{Kick1New}) and (\ref{Kick2New}). Thus, by using Eqs. (\ref{ExtensiveRule}), (\ref{1/m}), and (\ref{1/m^s}), we can obtain the upper bounds of $J^{(n)}$ and thereby $j_{\mathrm{max},n}$ for any $n$ in principle.


\providecommand{\noopsort}[1]{}\providecommand{\singleletter}[1]{#1}%


\begin{thebibliography}{48}
\bibitem [{Ord()}]{OrderParameter}%
  \BibitemOpen
  \href@noop {} {}\bibinfo {note} {The genuine order parameter is
  $e^{-iK_n}\sigma_i^z e^{iK_n}$, but it does not matter for the discussion
  since $e^{-iK_n}\sigma_i^z e^{iK_n}=\sigma_i^z+O(\lambda T)$.}\BibitemShut
  {Stop}%
\end{thebibliography}
\end{document}